\documentclass{article}
\usepackage[utf8]{inputenc}
\usepackage{amsmath}
\usepackage{amssymb}
\usepackage{hyperref}

\begin{document}
\title{Virtual beams and the Klein paradox for the Klein-Gordon equation}
\author{Alberto Molgado, Ociel Morales and Jos\'e A. Vallejo \\
{\normalsize Facultad de Ciencias\footnote{Lat. Av. Salvador Nava s/n. Col. Lomas,
CP 78290, San Luis Potos\'i (SLP) M\'exico.},
Universidad Aut\'onoma de San Luis Potos\'i}\\
{\normalsize and Dual CP Institute of High Energy, M\'exico}\\
{\footnotesize Emails: \texttt{molgado@fc.uaslp.mx,jvallejo@fc.uaslp.mx}}
}
\date{\today}
\maketitle

\begin{abstract}

Whenever we consider any relativistic 
quantum wave equation we are confronted with 
the Klein paradox, which asserts that incident particles 
will suffer a surplus of reflection when
dispersed by a discontinuous potential.
Following recent results on the Dirac equation,
we propose a solution to this paradox for the 
Klein-Gordon case by introducing
virtual beams in a natural well-posed generalization of the 
method of images in the theory of partial differential 
equations. Thus, our solution considers a global 
reflection coefficient obtained from the two contributions,
the reflected particles plus the incident virtual 
particles.  Despite its simplicity, this method allows a reasonable
understanding of the paradox within
the context of the quantum relativistic theory of 
particles (according to the original setup for 
the Klein paradox) and without resorting
to any quantum field theoretic issues.

\bigskip
\noindent\emph{Keywords}: Klein-Gordon equation; Klein paradox; Method of images.

\bigskip
\noindent PACS numbers: 03.65.Pm; 02.30.Jr.

\end{abstract}

\section{Introduction}

Back in the early days of quantum mechanics, following a suggestion of Debye,
Schr\"odinger tried to find an equation describing the behavior of the waves
introduced by De Broglie. His first attempt started with Einstein's equation
for the relativistic energy
\begin{equation}\label{einstein-eq}
E^2=p^2c^2+m^2c^4\,,
\end{equation}
which he expected to be satisfied by the quantum waves.
To this end, he observed that a
possible solution for the wave equation (deduced from Maxwell's equations for the components of the electromagnetic field)
\begin{equation}\label{wave-eq}
\frac{\partial^2 \Phi}{\partial x^2}-
\frac{1}{c^2}\frac{\partial^2 \Phi}{\partial t^2}=0\,,
\end{equation}
is a plane monochromatic wave
$$
\Phi(x,t)=\Phi_0 e^{i(kx-wt)}\,,
$$
where $k$ is the wave number and $w$ the frequency (for simplicity, we deal only with the one-dimensional case). Taking into account Einstein's relation
$E=\hbar w$, and De Broglie's one, $p=\hbar k$, he rewrote the expression for the plane waves arising in electromagnetic theory in terms of the energy and momentum, as
$$
\Phi(x,t)=\Phi_0 e^{i(px-Et)/\hbar}\,.
$$
Substituting back in \eqref{wave-eq}, of course we get the relativistic energy for
a photon (the quantum of the electromagnetic field)
$$
E^2=p^2c^2\,,
$$
but, as previously stated, Schr\"odinger wanted \eqref{einstein-eq} instead.
What he noticed, by a simple inspection, is that
this relation is precisely what results if we take as the wave equation
\begin{equation}\label{kg-eq}
\left(  
\frac{\partial^2 }{\partial x^2}-
\frac{1}{c^2}\frac{\partial^2}{\partial t^2}-\frac{m^2c^2}{\hbar^2}
\right)\Phi(x,t)=0\,.
\end{equation}
Schr\"odinger went on to study the Hidrogen atom with this equation, but soon
he found that it gave an incorrect spectrum, so he discarded it as a valid
quantum equation. After this, the equation was rediscovered by Pauli, Klein, 
Gordon and Fock, among others (Pauli called it ``the equation with many fathers'', see \cite{Enz02}), and
today \eqref{kg-eq} is widely known as the Klein-Gordon equation. It is a
quantum relativistic wave equation, used in the description of particles with
spin $0$.

As in the case of the Schr\"odinger equation (which he derived later starting from the non-relativistic expression for the energy), it is instructive to explore
the behavior of the solutions to the Klein-Gordon equation for some simple
potentials. Perhaps the simplest is the step potential,
\begin{equation}\label{step-eq}
V(x)=\begin{cases}
V,\mbox{ if }x\geq 0\,,\\[4pt]
0,\mbox{ if }x< 0\,.
\end{cases}
\end{equation}
When the energy of an incident beam of particles (from the left) is $E<V$, it
is well-known that for particles described by the Schr\"odinger equation there is some penetration in the region to the right of the barrier, expressed by the fact that there is a non vanishing transmission coefficient, but that density exponentially decays with distance.\\
What is surprising is that, as we shall see, when the step potential is included
in the Klein-Gordon equation the barrier can become transparent at high values
for $V$, and the (normalized) reflection coefficient is greater than $1$, even
when the energy of the incident beam is $E<V$. This counter-intuitive behavior is
called the Klein paradox. It is a common feature of quantum relativistic 
one-particle equations, and, indeed, it was first discussed for the Dirac
equation (see \cite{Kle29} and \cite{DC99}).

The Klein paradox is commonly solved for the case of the Dirac equation
within the framework of quantum field theory
(QFT) \cite{KSG04}: the strong potential to the right of the barrier excites the vacuum, creating electron-positron pairs, and acts attracting the positron states,
which couples
with the electrons outside the barrier with the same energy. Other solutions are
offered in \cite{DS07} and \cite{HR08}. However, there are
no available experiments where this phenomenon is considered, although recent
studies seem to point to bi-layer graphene as a possible setup \cite{KNG06}.
Anyway, as stated
in \cite{Alh11}, the main point is that the paradox arises in the context of a quantum
relativistic theory for one particle, so it would be desirable (to assess the self-consistency of the theory) to know if it can
be solved also without resorting to QFT. Indeed, we follow the ideas in that paper
(which deals with the Dirac equation) to show that this is the case for the
Klein-Gordon equation (there exists an earlier, different treatment in
\cite{Win59} based on the consideration of a finite non-zero width of the barrier,
and \cite{BRH98} reviews the usual pair creation solution in this setting). As in \cite{Alh11}, our main tool will be a suitable extension
of the method of images.

The method of images is described in any textbook dealing with electrostatics
as an efficient tool for studying the field created by charge distributions
involving media discontinuities. Its basic idea is to introduce virtual charges
inducing a field that compensates the one created by real charges, in such a
way that it satisfies the prescribed boundary condition on the media discontinuities.
Here, following \cite{Alh11}, we generalize it to show that if virtual beams, instead of virtual charges,
are introduced into the problem of the step potential for the Klein-Gordon equation, with suitable matching conditions, then the paradox is solved by
considering the total reflection and transmission coefficients (that is, the
coefficients corresponding to both the real and the virtual beams).

To make the paper relatively self-contained, we very briefly recall the basics of the 
method of images in section \ref{sec-prelim}, and reproduce in detail the computations
leading to the Klein paradox in section \ref{sec-Klein}. Section \ref{sec-sol}
contains the solution of the problem.

\section{Preliminaries: the method of images}\label{sec-prelim}

Let us begin by recalling that for any linear partial differential operator of
order $m$ with real-analytic coefficients on $\mathbb{R}^n$,
$P(\mathbf{x},D)$, and any non-characteristic, 
analytic regular submanifold $S\subset \mathbb{R}^n$, Holmgren's uniqueness theorem
(see \cite{Rau91}) guarantees the uniqueness of the solution to the Cauchy problem
$$
\begin{cases}
Pu=f \\[4pt]
D^\alpha u=D^\alpha g\mbox{ on }S\,,
\end{cases}
$$
for any multi-index $0\leq |\alpha|\leq m-1$, where $g$ is a given analytic function.
This is the case, for example, of the Laplace operator $\Delta =\nabla^2$ or
the Klein-Gordon one given in \eqref{kg-eq}.

Consider now the Poisson equation, defined on a connected domain 
$D\subset\mathbb{R}^n$ with regular boundary $S=\partial D$:
\begin{equation}\label{poisson-eq}
\nabla^2 u=f\,,
\end{equation}
where $f:D\to\mathbb{R}$ is a given function, physically representing, 
for instance, a distribution of electrostatic sources in empty space.
The fundamental solution for this
problem is a distribution (in the sense of L. Schwartz, see \cite{Sch50} or the more 
physics-oriented text \cite{Sch91}) $F(\mathbf{x}-\mathbf{x}')$ such 
that\footnote{Note that,
for a general differential operator, $F$ will not be a regular distribution, much less 
will have a functional dependence on $\mathbf{x},\mathbf{x}'\in\mathbb{R}^n$
through the difference $\mathbf{x}-\mathbf{x}'$. In this case, however, this happens 
because the Laplace operator has constant coefficients,
so translational invariance applies.}
\begin{equation}\label{green-eq}
\nabla^2 F(\mathbf{x}-\mathbf{x}')=\delta(\mathbf{x}-\mathbf{x}')\,,
\end{equation}
where $\delta$ is the Dirac (singular) distribution. Because of the property
$\delta \ast f=f=f\ast \delta$, it turns out that the solution to
\eqref{poisson-eq} is given by convolution with the source:
$$
u=F\ast f\,.
$$
The fundamental solution is not unique, of course. For instance,
the addition of any
harmonic distribution $h$ (such that $\nabla^2 h=0$) gives a new fundamental
solution $G=F+h$. The method of images provides a judicious choice of $h$ to
guarantee that prescribed boundary conditions on $S=\partial D$ are satisfied.\\
For the case of Dirichlet and Neumann problems, one wants that the solution
$G$ to satisfy $G(\mathbf{x}-\mathbf{x}')=0$ for any $\mathbf{x}\in S$,
and in this case it is called the 
Green function. Other conditions, such as asymptotic ones, can be imposed on $G$.\\
When the sources have support on a set $\Gamma$, given any
$\mathbf{x}'\in\Gamma$ we construct the associated Green function as $G=F+h$ where
\begin{equation}\label{nablah-eq}
\nabla^2h =\sum^N_{j=1}q_j\delta(\mathbf{x}-\mathbf{x}_j)\,,
\end{equation}
and $q_j$, $\mathbf{x}_j$ are, respectively, a
set of weights and positions chosen in such a way that
$G=F+h$ satisfy $G(\mathbf{x}-\mathbf{x}')=0$ for all $\mathbf{x}\in S=\partial D$.\\
What we want to stress at this point is the fact that the boundary $S$ provides
the conditions required to build the solution, and these conditions very often
can be deduced from symmetry conditions. We will return to this point later, in
Section \ref{sec-sol}.

The prototypical example where this method is applied is that of a harmonic
function in $\mathbb{R}^3$ with prescribed values on the plane $z=0$. Thus,
we look for a function $u(x,y,z)$ such that $\nabla^2u=0$ and
$u(x,y,0)=f(x,y,0)$ for a given function $f(x,y,0)$, which exists and is unique
because the Cauchy-Kowalevskaya and Holmgren theorems. Thus, if we consider the
problem of determining the field created by a single charge $q$ located at a distance
$d$ of an infinite, grounded conducting plane (which plays the r\^ole of the
surface $S$), we can arrange the coordinate system
so the conductor coincides with the $z=0$ plane and the charge is located at
$z=d$; then, the problem reduces itself to finding the location 
$\mathbf{x}_1$ of a single 
charge such that the total potential in the points $(x,y,0)$ vanishes. As stated above, 
symmetry considerations are an essential part of the method. In the present case,
the symmetry
of the problem suggests to take $\mathbf{x}_1=(0,0,-d)$ and $q_1=-q$ in 
\eqref{nablah-eq}. The Green function becomes the sum of the fundamental solution
of the Laplacian in $\mathbb{R}^3$ (see \cite{Sch91}), and the harmonic function
$h$ so found:
$$
G(\mathbf{x}-\mathbf{x}')=-\frac{1}{4\pi \Vert \mathbf{x}-\mathbf{x}'\Vert }+
h(\Vert \mathbf{x}-\mathbf{x}'\Vert)
=-\frac{1}{4\pi \Vert \mathbf{x}-\mathbf{x}'\Vert }+
\frac{1}{4\pi \Vert \mathbf{x}+\mathbf{x}'\Vert }\,,
$$
and the uniqueness
property assures that the solution for the potential in the region $z>0$ is the well 
known expression
$$
V(x,y,z)=\frac{q}{4\pi}\left(  
\frac{1}{\sqrt{x^2+y^2+(z-d)^2}}+\frac{1}{\sqrt{x^2+y^2+(z+d)^2}}
\right)\,.
$$

\section{The Klein paradox}\label{sec-Klein}

Let us simplify the notation by taking natural units ($\hbar =1=c$), so the
Klein-Gordon equation in three spatial dimension becomes
$$
\left( \frac{\partial}{\partial t^2}-\Delta+m^2\right)
\Phi(\mathbf{x},t)=0\,,
$$
or, in terms of the operators $\partial_\mu =(\partial/\partial t,\nabla)$ and
$\partial^\mu =\eta^{\mu\nu}\partial_\nu=(\partial/\partial t,-\nabla)$
(where $\eta_{\mu\nu}=\mathrm{diag}(1,-1,-1,-1)$ is Minkowski's metric),
$$
(\partial_\mu\partial^\mu +m^2)\Phi (x,t)=0\,.
$$
The relativistic probability current for the Klein-Gordon equation is defined as
the $4-$vector
$$
j^\mu =-\frac{1}{2mi}\left(\Phi^*\partial^\mu\Phi-\Phi\partial^\mu\Phi^* \right)\,.
$$
It is readily seen that $\partial_\mu j^\mu =0$, that is, the probability current is
conserved. The associated current density is given, as usual, by the temporal
component:
\begin{equation}\label{rho-eq}
\rho =j^0=-\frac{1}{2mi}\left(\Phi^*\partial^0\Phi-\Phi\partial^0\Phi^* \right)=
-\frac{1}{2mi}\left(\Phi^*\partial_t\Phi-\Phi\partial_t\Phi^* \right)\,,
\end{equation}
so, from the conservation of $j^\mu$ we get the continuity equation
$$
\partial_\mu j^\mu =\frac{\partial \rho}{\partial t}+\nabla \mathbf{j}=0\,,
$$
where $j^\mu =(\rho,\mathbf{j})$. Notice that, for a free plane wave solution
of definite $3-$momentum $\mathbf{p}$,
$\Phi(\mathbf{x},t)=e^{-i(Et-\mathbf{p}\cdot\mathbf{x})}$, the density is given
by
$$
\rho(\mathbf{x},t)=\frac{E}{m}\,,
$$
and the fact that we are dealing with relativistic particles, where the energy
and momentum are related through \eqref{einstein-eq}, implies that $E$ can take any of
the values $E=\pm \sqrt{\Vert\mathbf{p}\Vert^2 +m^2}$, so $\rho$ is not 
positive-definite and cannot be interpreted as a probability density (that is the
reason for calling it a current density instead). This appearance of negative energies lies at the heart of the Klein paradox, as we will see in what follows.

When including interaction with an electromagnetic field, use must be made of
the minimal coupling, replacing $p_\mu$ by $p_\mu -A_\mu$ (natural units), where
$A_\mu =(V,\mathbf{A})$ is the electromagnetic potential. In this case, it is
easy to see that the density associated to the the conserved current is
\begin{equation}\label{rho2-eq}
\rho =-\frac{1}{2mi}\left(\Phi^*\partial_t\Phi-\Phi\partial_t\Phi^* \right)
-\frac{1}{m}V\Phi^*\Phi\,,
\end{equation}
instead of \eqref{rho-eq}.

Consider now a beam of particles in one spatial dimension, of positive unit charge, 
described by the Klein-Gordon equation in the presence of a step (electric) potential 
\eqref{step-eq}. If the particles fall on the barrier from the left, with an energy $E$ 
satisfying $0<E<V$, by defining the effective momenta
\begin{equation}\label{mom-eq}
k_1=+\sqrt{E^2-m^2}\quad ,\quad k_2=+\sqrt{(V-E)^2-m^2}\,,
\end{equation}
it is readily found that the solution with positive energies is given by
\begin{equation}\label{phi-eq}
\Phi(x,t)=\begin{cases}
e^{-iEt}u_I(x),\quad x<0\\[4pt]
e^{-i(V-E)t}u_{II}(x),\quad x\geq 0\,,
\end{cases}
\end{equation}
with the stationary functions
\begin{equation}
\begin{cases}\label{u-eq}
u_I=e^{ik_1 x}+Ae^{-ik_1 x},\quad x<0 \\[4pt]
u_{II}=Ce^{ik_2 x}+Be^{-ik_2 x},\quad x\geq 0\,,
\end{cases}
\end{equation}
and where we have normalized the incident beam, so the coefficient of the fraction that
propagates to the right in the region $x<0$ is $1$. Notice that this solution is
a superposition of beams propagating to the right and to the left in both regions,
$x<0$ (region I), and $x\geq 0$ (region II). Our identification of the incident beam
is based upon the fact that the current density in region I is given by \eqref{rho-eq},
$$
\rho_I = -\frac{1}{2mi}\left(\Phi^*\partial_t\Phi-\Phi\partial_t\Phi^* \right)
=\frac{i}{2m}(-2iEu^*_Iu_I)=\frac{E}{m}|u_I|^2\,,
$$
so it has the same sign as the incident charge, while in the region II it has the
opposite sign, because $V>E$ in \eqref{rho2-eq},
$$
\rho_{II} = \frac{i}{2m}(-2iEu^*_{II}u_{II})-
\frac{V}{m}u^*_{II}u_{II}=\frac{E-V}{m}|u_{II}|^2\,.
$$
Thus, in the expression for $u_{II}$ we find that the term $Ce^{ik_2x}$ has an
associated current running from left to right, with a negative current density
$\rho_{II}=|C|^2(E-V)/m$; equivalently, we could say that it describes a current
of positive charges running from right to left. Analogously, the term
$Be^{-ik_2x}$ has an associated current running from right to left with negative
current density, so it can be interpreted as a positive current from left to right.
As the only charges present are positive, and there are no sources to the right of
the barrier, we must take $C=0$ in \eqref{u-eq}, arriving at the proposed solution
\begin{equation}\label{stat-sol}
\begin{cases}
u_I=e^{ik_1 x}+Ae^{-ik_1 x},\quad x<0 \\[4pt]
u_{II}=Be^{-ik_2 x},\quad x\geq 0\,.
\end{cases}
\end{equation}
As a technical remark, let us note that a straightforward computation with the
wavefunction \eqref{phi-eq} (where the stationary solutions are given by
\eqref{stat-sol}) gives
$$
(\partial_\mu\partial^\mu +m^2)\Phi(x,t)=
\begin{cases}
(m^2-E^2+k^2_1)\Phi(x,t),\quad x<0 \\[4pt]
(m^2-(V-E)^2+k^2_2)\Phi(x,t),\quad x\geq 0\,,
\end{cases}
$$
so, in order to be a solution of the Klein-Gordon equation, $\Phi(x,t)$ must have
its support contained in the hyperboloids $m^2-E^2+k^2_1=0$, for $x<0$, and
$m^2-(V-E)^2+k^2_1=0$, for $x\geq 0$. These on-shell conditions are automatically
satisfied because of the definitions \eqref{mom-eq} of the effective momenta.

In this way, we have identified the incident, reflected and transmitted beams,
$u_{i}$, $u_r$, and $u_t$, respectively given as
$u_i=e^{ik_1x}$, $u_r=Ae^{-ik_1x}$, and $u_t=Be^{-ik_2x}$. Now, we can compute the
corresponding incident, reflected, and transmitted currents:
\begin{align*}
j_i=& \frac{1}{2mi}\left( 
u^*_i\frac{\partial}{\partial x}u_i-u_i\frac{\partial}{\partial x}u^*_i
\right)=\frac{k_1}{m} \\
j_r=& \frac{1}{2mi}\left( 
u^*_r\frac{\partial}{\partial x}u_r-u_r\frac{\partial}{\partial x}u^*_r
\right)=-\frac{|A|^2k_1}{m} \\
j_t=& \frac{1}{2mi}\left( 
u^*_t\frac{\partial}{\partial x}u_t-u_t\frac{\partial}{\partial x}u^*_t
\right)=-\frac{|B|^2k_2}{m}\,.
\end{align*}
Then, we get for the reflection coefficient the expression
$$
R=\left|\frac{j_r}{j_i} \right|=|A|^2\,,
$$
while the transmission coefficient is
$$
T=\frac{j_t}{j_i}=-\frac{k_2}{k_1}|B|^2\,.
$$
Imposing the continuity conditions at the barrier for the wave function and
its derivative, $u_I(0)=u_{II}(0)$, and $u'_I(0)=u'_{II}(0)$, we get a system
of equations for $A$, $B$ whose solution is
$$
A=\frac{k_1+k_2}{k_1-k_2}\,,\quad B=\frac{2k_1}{k_1-k_2}\,,
$$
so we can write, in terms of the effective momenta,
\begin{equation}\label{rt-eq}
R=\left( \frac{k_1+k_2}{k_1-k_2} \right)^2\,,\quad
T=-\frac{k_2}{k_1}\left( \frac{2k_1}{k_1-k_2} \right)^2\,.
\end{equation}
Thus, the reflection and transmission coefficients make a pretense of respecting the
conservation of matter, as a straightforward computations shows that
$$
R+T=1\,.
$$
However, a look at \eqref{rt-eq} reveals that $T$ is negative, an already shocking 
fact, but even more so since it implies $R>1$, meaning that there are more reflected 
than incident particles. This is the Klein paradox.

\section{Virtual beams and resolution of the paradox}\label{sec-sol}

We will exploit the similarity between the geometric setup of the Klein-Gordon paradox 
and that of the method of images. 
In both cases we have a surface of discontinuity which
determines some conditions to be satisfied by the solution to the problem although, of 
course, there is a fundamental difference: The method of images is applied to a static 
charge configuration, while in the Klein-Gordon case we have density currents; but let 
us focus on the formal similarities as a means to obtain an \emph{ansatz} for the 
structure of the solution.

Following the analogy with the method of images, we look for a way to compensate the
excess in the reflection coefficient $R>1$, found in the preceding section, by 
considering it as a measure of the flux of particles in a beam through the 
discontinuity surface determined by the barrier. Thus, we will construct virtual
beams out from symmetry considerations in such a way that the net flux through the 
barrier, from left to right, gives $R<1$. Similarly, the net transmission coefficient
must be $T<1$ while their sum satisfies $R+T=1$. As it is the case with the virtual 
image introduced in the example of Section \ref{sec-prelim}, these virtual beams are 
not observable, so they can be safely incorporated to the solution.

Now, recall that the Green solution given by the method of images has the structure
$G=F+h$, where $F$ is a fundamental solution and $h$ is chosen so $G$ satisfies the
prescribed boundary conditions. Let us choose as the analog of $F$ the wavefunction 
$\Phi(x,t)$
given by \eqref{phi-eq} with \eqref{stat-sol}.
Due to the symmetry of the problem, as the function $h$ we choose the wavefunction
describing a beam of particles incident on the barrier from right to left,
whose stationary part is given by
$$
w(x)=
\begin{cases}
w_I=e^{ik_2x}+Ce^{-ik_2x}\,,\quad x\geq 0 \\[4pt]
w_{II}=De^{-ik_1x}\,,\quad x<0\,.
\end{cases}
$$
It is straightforward to compute, for the function $\Psi(x,t)=e^{-iEt}w_I$ on $x<0$,
and $\Psi(x,t)=e^{-i(V-E)t}w_{II}$ on $x\geq 0$, that
$$
(\partial_\mu\partial^\mu +m^2)\Psi(x,t)=
\begin{cases}
(m^2-E^2+k^2_1)\Psi(x,t),\quad x<0 \\[4pt]
(m^2-(V-E)^2+k^2_2)\Psi(x,t),\quad x\geq 0\,,
\end{cases}
$$
so it is supported on the same subset as $\Phi(x,t)$. Thus, on-shell we have two
solutions to the Klein-Gordon equation and by linearity we can form their
superposition. Actually, we only need to work with the
stationary solutions, so we will forget about the time dependence in what follows.

Let us compute the reflection and transmission coefficients for the solution $w$.
A reasoning completely analogous to that developed in the preceding section leads
to the incident, reflected and transmitted beams $w_i$, $w_r$, and $w_t$ given by
$w_i=e^{ik_2x}$, $w_r=Ce^{-ik_2x}$, and $w_t=De^{-ik_1x}$, respectively. Then,
we can compute the corresponding currents (where we have added a superscript to
make clear the solution used),
\begin{align*}
j^w_i=& \frac{1}{2mi}\left( 
w^*_i\frac{\partial}{\partial x}w_i-w_i\frac{\partial}{\partial x}w^*_i
\right)=\frac{k_2}{m} \\
j^w_r=& \frac{1}{2mi}\left( 
w^*_r\frac{\partial}{\partial x}w_r-w_r\frac{\partial}{\partial x}w^*_r
\right)=-\frac{|C|^2k_2}{m} \\
j^w_t=& \frac{1}{2mi}\left( 
w^*_t\frac{\partial}{\partial x}w_t-w_t\frac{\partial}{\partial x}w^*_t
\right)=-\frac{|D|^2k_1}{m}\,.
\end{align*}
For this virtual beam, the expressions of the reflection and transmission coefficients
are
$$
R^w=\left|\frac{j^w_r}{j^w_i} \right|=|C|^2\,,\quad
T^w=\frac{j^w_t}{j^w_i}=-\frac{k_1}{k_2}|D|^2\,.
$$
A look at these formulas reveals a symmetric behavior with respect to the wavefunction
that represents the beam running from left to right. Something more can be said if we
consider the continuity conditions for $w(x)$ and $w'(x)$ at the barrier. Indeed, these
conditions lead to a system of equations for $C$, $D$ with solution
$$
C=\frac{k_1+k_2}{k_2-k_1}\,,\quad D=\frac{2k_2}{k_2-k_1}\,,
$$
leading to the following expressions for $R^w$, $T^w$, in terms of the effective 
momenta,
\begin{equation}\label{rtw-eq}
R^w=\left(\frac{k_1+k_2}{k_2-k_1} \right)^2\,,\quad 
T^w=-\frac{k_1}{k_2}\frac{4k^2_2}{(k_2-k_1)^2}\,.
\end{equation}
It is readily seen that, again, $R^w>1$ and $T^w<0$. Moreover, we have the fake
conservation relation
$$
R^w+T^w=1\,.
$$
However, the superposition $u(x)+w(x)$ solves all the problems. It has associated
a global reflection coefficient obtained by considering the fraction of particles
(whatever their charges) that after the dispersion lie at the left of the barrier.
This fraction come from two sources. On the one hand, we have the reflected particles
that were incident form the left. On the other hand, there are the transmitted
virtual particles that were incident from the right. This is analogous as the 
computation of the potential on the boundary surface in the method of images, where
we superimpose the potential of the real and virtual charges.

Therefore, the global reflection coefficient $R_G$ can be computed by adding the 
reflection coefficient for the original beam, $R^u$ in \eqref{rt-eq} (putting the $u$ 
superscript for clarity), and the transmission coefficient for the virtual beam, 
$T^w$ in \eqref{rtw-eq}. We then obtain, after some simple algebraic manipulations,
$$
R_G=R^u+T^w=\left( \frac{k_1+k_2}{k_1-k_2} \right)^2
-\frac{k_1}{k_2}\frac{4k^2_2}{(k_2-k_1)^2}=1\,.
$$
This result can be interpreted by saying that the superposition $u(x)+w(x)$ leads to
a total global reflection of the particles to the left, what is to be expected since
the incident energy is $E<V$, thus solving the paradox.

\section{Conclusions}

The Klein paradox appears for any relativistic quantum wave equation when studying
the dispersion by a discontinuous potential. Following the treatment in
\cite{Alh11} for the Dirac equation, we have shown that in the case of the Klein-Gordon 
equation the excess in the reflection coefficient can be explained by the introduction
of virtual beams, analog to the introduction of virtual charges in the well-known 
method of images of electrostatics. A detailed mathematical analysis of this method 
allows a reasonable understanding of the paradox within the context of the quantum
relativistic theory of one particle, without resorting to quantum field theoretic
ideas.

\section{Acknowledgements}
A. Molgado thanks the financial support from CONACyT--M\'exico under project number 
CB--2014--243433. 
Jos\'e A. Vallejo was partially supported by a project CONACyT--M\'exico
CB-2012--179115.

\end{document}